\documentclass[12 pt, amsfonts, amssymb,color]{article}

\usepackage{amsmath,amssymb,amsfonts,latexsym,float,graphics,epsfig}
\usepackage{subcaption}
\usepackage{color}

\begin{document}
%<<<<<<<<<<< enumeration of eqns section wise>>>>>>>>>>>>>>>>>>>
\renewcommand\theequation{\arabic{section}.\arabic{equation}}
\catcode`@=11 \@addtoreset{equation}{section}
%<<<<<<<<<<<<<<<<<<<<<<<<<<<<<<<<<>>>>>>>>>>>>>>>>>>>>>>>>>>>>>>>>>
\newtheorem{axiom}{Definition}[section]
\newtheorem{theorem}{Theorem}[section]
\newtheorem{axiom2}{Example}[section]
\newtheorem{lem}{Lemma}[section]
\newtheorem{claim}{Claim}[section]
\newtheorem{prop}{Proposition}[section]
\newtheorem{cor}{Corollary}[section]
\newcommand{\be}{\begin{equation}}
\newcommand{\ee}{\end{equation}}
\newcommand{\equal}{\!\!\!&=&\!\!\!}
\newcommand{\rd}{\partial}
\newcommand{\g}{\hat {\cal G}}
\newcommand{\bo}{\bigodot}
\newcommand{\res}{\mathop{\mbox{\rm res}}}
\newcommand{\diag}{\mathop{\mbox{\rm diag}}}
\newcommand{\Tr}{\mathop{\mbox{\rm Tr}}}
\newcommand{\const}{\mbox{\rm const.}\;}
\newcommand{\cA}{{\cal A}}
\newcommand{\bA}{{\bf A}}
\newcommand{\Abar}{{\bar{A}}}
\newcommand{\cAbar}{{\bar{\cA}}}
\newcommand{\bAbar}{{\bar{\bA}}}
\newcommand{\cB}{{\cal B}}
\newcommand{\bB}{{\bf B}}
\newcommand{\Bbar}{{\bar{B}}}
\newcommand{\cBbar}{{\bar{\cB}}}
\newcommand{\bBbar}{{\bar{\bB}}}
\newcommand{\bC}{{\bf C}}
\newcommand{\cbar}{{\bar{c}}}
\newcommand{\Cbar}{{\bar{C}}}
\newcommand{\Hbar}{{\bar{H}}}
\newcommand{\cL}{{\cal L}}
\newcommand{\bL}{{\bf L}}
\newcommand{\Lbar}{{\bar{L}}}
\newcommand{\cLbar}{{\bar{\cL}}}
\newcommand{\bLbar}{{\bar{\bL}}}
\newcommand{\cM}{{\cal M}}
\newcommand{\bM}{{\bf M}}
\newcommand{\Mbar}{{\bar{M}}}
\newcommand{\cMbar}{{\bar{\cM}}}
\newcommand{\bMbar}{{\bar{\bM}}}
\newcommand{\cP}{{\cal P}}
\newcommand{\cQ}{{\cal Q}}
\newcommand{\bU}{{\bf U}}
\newcommand{\bR}{{\bf R}}
\newcommand{\cW}{{\cal W}}
\newcommand{\bW}{{\bf W}}
\newcommand{\bZ}{{\bf Z}}
\newcommand{\Wbar}{{\bar{W}}}
\newcommand{\Xbar}{{\bar{X}}}
\newcommand{\cWbar}{{\bar{\cW}}}
\newcommand{\bWbar}{{\bar{\bW}}}
\newcommand{\abar}{{\bar{a}}}
\newcommand{\nbar}{{\bar{n}}}
\newcommand{\pbar}{{\bar{p}}}
\newcommand{\tbar}{{\bar{t}}}
\newcommand{\ubar}{{\bar{u}}}
\newcommand{\utilde}{\tilde{u}}
\newcommand{\vbar}{{\bar{v}}}
\newcommand{\wbar}{{\bar{w}}}
\newcommand{\phibar}{{\bar{\phi}}}
\newcommand{\Psibar}{{\bar{\Psi}}}
\newcommand{\bLambda}{{\bf \Lambda}}
\newcommand{\bDelta}{{\bf \Delta}}
\newcommand{\p}{\partial}
\newcommand{\om}{{\Omega \cal G}}
\newcommand{\ID}{{\mathbb{D}}}
\newcommand{\pr}{{\prime}}
\newcommand{\prr}{{\prime\prime}}
\newcommand{\prrr}{{\prime\prime\prime}}

\title{Relativistic formulation of curl force, relativistic Kapitza equation and trapping}

\author{Partha Guha\footnote{E-mail: partha.guha@ku.ac.ae}\\
Department of Mathematics\\
Khalifa University of Science and Technology \\ Main Campus, P.O. Box -127788, Abu Dhabi\\
United Arab Emirates\\
\and
Sudip Garai\footnote{E-mail: sudip.dhwu@gmail.com}\\
Department of Physics \\
Diamond Harbour Women's University \\
D. H. Road, Sarisha, West-Bengal 743368, India\\
}

\date{\today}

\maketitle

\smallskip

\begin{abstract}
In this present communication the relativistic formulation of the curl forces with saddle potentials has been performed. In particular, we formulated the relativistic version of the Kapitza equation. The dynamics and trapping phenomena of this equation have been studied both theoretically and numerically. The numerical results show interesting characteristics of the charged particles associated with the particle trapping and escaping in the relativistic domain. In addition, the relativistic generalization of the Kapitza equation associated with the monkey saddle has also been discussed.
\end{abstract}

\smallskip


\smallskip

\noindent


\smallskip


\section{Introduction}

A Newtonian dynamics associated with a non-zero curl force has been a topic of interest to many of the researchers round the globe in the recent past. These forces are known as curl forces which has been introduced to the research community by Berry and Shukla\cite{BS466,BS3}. These forces, in general, are velocity independent in order to satisfy a non-zero curl \emph{viz.} force $F$ depends only on position ${\bf r}$ and is independent of velocity ${\bf v}$. The interesting fact is that these curl forces can not be derivable from a scalar potential. The direct applications about the understanding of the system characteristics associated with these curl forces span to many compulsive fields like optics, laser physics, noncentral forces, anisotropic Kepler problem etc.\cite{BS2,CNV,AML,SS,Gutz,Devaney,GCGP,Guha1,GaraiGuha,GaraiGuha2}. The generalization of these curl forces have been made with the introduction of dissipation factor present in the system and gyroscopic terms. The application to this directs us to ion trapping, two level atomic state, gain-loss mechanisms etc.

\bigskip

The curl force preserves the volume in phase space domain \emph{viz.} $({\bf r}, {\bf v})$ without any attractors and this can be visualized as follows. Let us take the dynamical equation for a unit mass as $\ddot{\bf r} = F({\bf r})$ with a non zero curl i.e. $\nabla \times F({\bf r}) \neq 0$. Henceforth one can observe that
$$
\nabla_{\bf r}{\bf v} + \nabla_{\bf v}\dot{\bf v} = \nabla_{\bf r}{\bf v} + \nabla_{\bf v}F({\bf r}) = 0,
\qquad {\bf v} = \dot{\bf r}.
$$
Although, in general, the dynamics associated with the curl forces are non-dissipative. Therefore in presence of dissipation the phase space volume is not preserved under the curl force, i.e. $\nabla \times F({\bf r}) \neq 0$ and $\nabla_{\bf r}{\bf v} + \nabla_{\bf v}F({\bf r}) \neq 0$.

\bigskip

The curl forces are generated by the dynamical motion associated with saddle potentials. A particle remains stable to its equilibrium if the surface rotates around the vertical axis sufficiently fast. To visualize let us take a saddle force field as $ \big(F_x = x, F_y = -y\big)$. We now covert this to a time-dependent one by implementing a rotation to each of the vector counterclockwise with angular velocity $2\omega$. This describes the dynamical motion of a particle under this force field with unit mass. From the Earnshaw’s theorem we know that a static potential can not trap a charged particle because any static potential, which satisfies the Laplace’s law, lacks a potential minimal. Thus the trapping of a charged particle can be achieved by rotating the saddle potential around a suitable axis with a suitable angular frequency. This was the famous idea of Wolfgang Paul\cite{Paul}. The basic idea of Paul was to stabilize the saddle by means of a `vibrating' electrostatic field by analogy with so called \emph{Stephenson-Kapitsa}\cite{Stephenson,KapitsaPen} pendulum. On the other hand, in the Penning trap, magnetic fields are used for charged particle's trapping\cite{Penning1,Penning2,Penning3}. In recent times, the study regarding the stabilization and other geometric properties of these highly interesting curl forces and the dynamics of nonlinear Hamiltonian curl forces have been reported\cite{GCGu,KL1,KL2}.

\bigskip

The dynamics of the curl forces are related to the theory of Kapitsa-Merkin nonconservative positional forces. The linearized dynamics of a rotating shaft formulated by Kapitsa\cite{Kapitsa} is given by
\be \label{Kapitsa}
\ddot{x} + ay + bx = 0, \qquad \ddot{y} - ax + by = 0.
\ee
The corresponding characteristics equation reflects that, if we add a non-zero nonconservative curl force ( i.e. $ a \neq 0$) to a stable system associated with a stable potential then it becomes unstable. This is associated to the Merkin's result\cite{Merkin,Merkin1}, which states that ``the introduction of nonconservative linear forces into a system with a stable potential and with equal frequencies destroys the stability regardless of the form of nonlinear terms''. It is worth mentioning that the positional force, i.e. the terms $ay$ and $-ax$ are proportional to $\omega^2$, where $\omega$ is the rotation rate of the shaft. It is also possible to derive Eq.(\ref{Kapitsa}) associated with a saddle potential via \emph{Euler-Lagrange} method. It is very straightforward to check that the  Lagrangian \be L = \frac{1}{2}(\dot{x}^{2} - \dot{y}^{2}) - \frac{1}{2}a(x^2 - y^2) - bxy, \ee yields Eq.(\ref{Kapitsa}). A symmetric saddle surface can be described  by ${\tilde U} = a(x^2 - y^2)$, where $a$ is a geometrical parameter that specifies the curvature of the saddle. It is note worthy to mention that an ion trap potential is a rotating saddle surface on which a ball can be trapped. The pondermotive potential of an ion trap is that of a saddle in $2D$ but the time evolution of the potential is one that flaps up and down. The potential of the modified rotating system is $${\Bbb U}(x,y,t) = \frac{1}{2}(x^2 - y^2)\cos(2\omega t) -xy \sin(2\omega t), $$ where $\omega$ is the angular drive frequency of the spinning saddle.

\bigskip

The relativistic extension of the nonlinear dynamical equations are not a straightforward job due to the intrinsic extra nonlinearity triggered by the presence of the Lorentz factor\cite{Harvey,LlibreMak,LiLiu}. The relativistic harmonic oscillator is a topic of immense interest,
although widely discussed, still contains elements of interest. Here the Newtonian kinetic energy is replaced by its special relativistic counterpart.
This version of the relativistic harmonic oscillator model has recently been probed experimentally also\cite{Fujiwara}. Recently, Haas\cite{Haas} generalized Ermakov systems towards the special relativity domain. In fact relativistic nonlinear dynamics is an open area of research and deserves
more attention and intense studies. In this present work we study the relativistic analog of curl force  and explore the relativistic Kapitza equation (RKE) and trapping phenomena. One must note that the RKE is the prototypical example of a relativistic curl force.

\bigskip

The paper is organized as follows: In Section $2$, we recapitulate, briefly, the construction of the linear curl forces by following Berry and Shukla. We also introduce the Kapitza equation and dynamics related to the higher saddles there. Section $3$ is dedicated to the relativistic extension of the curl forces and Kapitza equation. We also describe the trapping phenomena in that section. Lastly, in Section $4$, we summarize our findings in the relativistic domain with the comparison to the non-relativistic results along with the possible application towards realistic scenarios.

\section{Preliminaries: linear curl forces and Hamiltonization}

The central fact about curl force theory is that only a small subset of all curl forces are Hamiltonian cases and they find nice applications also. But it is the non-Hamiltonian curl forces that are really distinctive and exhibit various new features of dynamics.

\bigskip

Let us summarise the classical theory of linear curl forces\cite{BS466,BS3,GCGP}. We recall from Ref.\cite{BS466} that the form of the Hamiltonian in the case of curl forces is given by
\be\label{E1}
H=\frac{1}{2}\alpha p_x^2 +\beta p_xp_y +\frac{1}{2}\gamma p_y^2
+U(x,y),
\ee
where the potential is defined by
\be\label{pot} U(x,y) = \frac{1}{2}ax^2 + bxy + \frac{1}{2}cy^2. \ee
The first set of Hamiltonian equations then gives
\be
\dot{x} = \alpha p_x + \beta p_y, \qquad \dot{y} = \beta p_x + \gamma p_y.
\ee
We can then get the forces from the second set of Hamiltonian equations as
\be
a_x = \ddot{x} = \alpha \dot{p_x} + \beta \dot{p_y} = -\alpha \frac{\partial U(x,y)}{\partial x} - \beta
\frac{\partial U(x,y)}{\partial y},
\ee
\be
a_y = \ddot{y} = \beta \dot{p_x} + \gamma \dot{p_y} = -\beta \frac{\partial U(x,y)}{\partial x} - \gamma
\frac{\partial U(x,y)}{\partial y}.
\ee
The curl is
\be \nabla \times {\bf F} = (\alpha - \gamma)\frac{\partial^2 U(x,y)}{\partial x \partial y}{\bf i}
+ \beta \big( \frac{\partial^2 U(x,y)}{\partial x^2} - \frac{\partial^2 U(x,y)}{\partial y^2}\big){\bf j},
\ee
which yields
$ \nabla \times {\bf F} = (\alpha - \gamma)b{\bf i}
 + \beta(c-a){\bf j}.$
To ensure that $\nabla\times F\ne 0$, it is necessary to take $\beta=0$  and $\alpha=1=-\gamma$. Also the another choice we have is $\beta \neq 0$ and $ c = -a$. Henceforth the Hamiltonian turns out to be
\be\label{E1a} H=\frac{1}{2}(p_x^2-p_y^2) + \frac{1}{2}(x^2-y^2) + bxy,\ee
which is a Hamiltonian of the Kapitza equation.

\subsection{Rotating saddle and Kapitza equation related to higher-saddle}

Consider the motion of a particle in the rotating saddle potential in the
plane
\be\label{First}
U({\bf x},t) = U_0(R^{-1}{\bf x}), \qquad U_0 = \frac{1}{2}(x^2 - y^2), \qquad {\bf x} = (x,y);
\ee
where $$ R = R(\Omega t) = \left({\begin{array}{cc}
                     \cos(\omega t) & \sin(\omega t) \\
                     -\sin(\omega t) & \cos(\omega t) \\
                       \end{array} }\right). $$
The Lagrangian of the rotating saddle potential is given by
\be
L_{rotsaddle} = \frac{1}{2}(\dot{x}^{2} - \dot{y}^{2}) - \frac{\Lambda}{2}
\Big( \big(x^2 - y^2 \big)\cos(2\omega t)
 + 2xy \sin(2\omega t) \Big), \,\,\,\,\, \Lambda = 2\frac{mgh_0}{r_{0}^{2}},
\ee
and the corresponding equations of motion are
\be
\ddot{x} + \Lambda x\cos(2\omega t) +  \Lambda y \sin(2\omega t) = 0, \qquad
\ddot{y} + \Lambda y\cos(2\omega t) - \Lambda x \sin(2\omega t) = 0.
\ee

\subsubsection{Higher saddle and generalized Kapitza equation}

We can generalized rotating shaft equation formulated by Kapitza using higher-order saddles. The Eq.(\ref{Kapitsa}) can be manufactured from the Euler-Lagrange equation using simple saddle potential $g_1(x,y) = x^2 - y^2$ and the rotated version of the same surface, $g_{1}^{R}(x,y) = xy$. In other words, latter one is obtained from the rotation of the saddle by $90$ degree. Let $z = x + iy$, then $x^2 - y^2 = Re(z^2)$ and $2xy = Im(z^2)$. We follow this guideline to construct equations of rotating shaft associated to monkey saddle
\be g_2(x,y) = x^3 - 3xy^2 = \hbox{ Re } (z^3) \ee and its rotated version \be g_{2}^{R}(x,y) = 3x^2y - y^3 = \hbox{ Im }
(z^3).\ee
The generalized rotating shaft pair of equations associated to degree three potential or monkey saddle potential is given by
$$ U_3 = \frac{ 1}{3} \big( k_1 g_2(x,y) + k_2g_{2}^{R}(x,y) \big) = \frac{ 1}{3} \big( k_1 (x^3 - 3xy^2)
+ k_2(3x^2y - y^3 \big). $$
The corresponding equations of motion are given by
\be \ddot{x} + k_1(x^2 - y^2) + 2k_2xy = 0, \quad \ddot{y} + 2k_1xy - k_2(x^2 - y^2) = 0. \label{rse1} \ee

\section{Relativistic curl force}

The relativistic Lagrangian in presence of scalar potential fields is given by
\be
L = -mc^2\sqrt{ 1 - 2\frac{\cal L}{mc^2}}, \quad {\cal L} = \frac{m|{\bf v}|^2}{2} - U(x).
\ee
One way to describe the non-relativistic approximation is to apply the limits: $U(x) << mc^2$; $v << c$,
$$
L = -mc^2\sqrt{ \big(1 - \frac{|{\bf v}|^2}{c^2}\big) + \frac{2U}{mc^2} }
= -mc^2\sqrt{ 1 - \frac{|{\bf v}|^2}{c^2}} \sqrt{1 + \frac{2U}{mc^2}\big(1 - \frac{|{\bf v}|^2}{c^2}\big)^{-1}}
$$
$$
= -mc^2 \gamma^{-1} \sqrt{1 + \frac{2U}{mc^2}\gamma^2}, \quad
\gamma = \big(1 - \frac{|{\bf v}|^2}{c^2}\big)^{-\frac{1}{2}}.
$$
Expanding the expression within square-root binomially up to second order gives us
\be
\tilde{L}  = -mc^2\gamma^{-1} \big( 1 + \frac{U}{mc^2}\gamma^2 \big) = -mc^2\gamma^{-1} - U\gamma.
\ee
Although in \emph{Goldstein}'s book the semi-relativistic Lagrangian is given by
\be L_{sr} -mc^2\gamma^{-1} - U. \ee
If we assume
$$
\frac{|{\bf v}|}{c} = \alpha << 1, \Rightarrow U\gamma \sim U + \frac{U}{2}\alpha^2,
$$
for practical purpose; we can drop the second term for small value of $\alpha$ which immediately yields the semi-relativistic Lagrangian given in \emph{Goldstein}\cite{Goldstein}.

\subsection{Relativistic Kapitza equation}

The relativistic Lagrangian of the curl force is defined as
\be
L = - \frac{c^2}{\Gamma} - U(x,y), \quad \hbox{ with } \quad
\Gamma = \big( 1 - \frac{\dot{x}^2}{c^2} +  \frac{\dot{y}^2}{c^2} \big)^{- \frac{1}{2}}.
\ee
Here the saddle potential is assumed as
\be
U(x,y) = \frac{1}{2}k(x^2 - y^2) + bxy.
\ee
\begin{lem}
The Euler-Lagrange equation of the relativistic Lagrangian $L = - \frac{c^2}{\Gamma} - \frac{1}{2}k(x^2 - y^2) - bxy$
yields
	\be\label{eqx}
	\ddot{x}( 1 + \frac{\dot{y}^2}{c^2}) - \frac{\dot{x}\dot{y}}{c^2}\ddot{y} = - \frac{kx}{\Gamma^3} - \frac{by}{\Gamma^3},
	\ee
\be\label{eqy}
	\ddot{y}( 1 - \frac{\dot{x}^2}{c^2}) + \frac{\dot{x}\dot{y}}{c^2}\ddot{x} =  -\frac{ky}{\Gamma^3} + \frac{bx}{\Gamma^3},
	\ee
where $\Gamma = ( 1 - \frac{\dot{x}^2}{c^2} +  \frac{\dot{y}^2}{c^2})^{- \frac{1}{2}}$.
\end{lem}
{\bf Proof} It is straightforward to check that
$$
\frac{\partial L}{\partial \dot{x}} = \dot{x}( 1 - \frac{\dot{x}^2}{c^2} +  \frac{\dot{y}^2}{c^2})^{- \frac{1}{2}},
\quad \frac{d}{dt}(\frac{\partial L}{\partial \dot{x}}) = \big[\ddot{x} (  1 + \frac{\dot{y}^2}{c^2}) -
\frac{\dot{x}\dot{y}}{c^2}\ddot{y}]( 1 - \frac{\dot{x}^2}{c^2} +  \frac{\dot{y}^2}{c^2})^{- \frac{3}{2}},
$$
$$
\frac{\partial L}{\partial \dot{y}} = -\dot{y}( 1 - \frac{\dot{x}^2}{c^2} +  \frac{\dot{y}^2}{c^2})^{- \frac{1}{2}},
\quad  \frac{d}{dt}(\frac{\partial L}{\partial \dot{y}}) = \big[-\ddot{y} (  1 - \frac{\dot{x}^2}{c^2}) -
\frac{\dot{x}\dot{y}}{c^2}\ddot{x}]( 1 - \frac{\dot{x}^2}{c^2} +  \frac{\dot{y}^2}{c^2})^{- \frac{3}{2}}.
$$
Then using the derivative of $\frac{\partial L}{\partial x}$ and $\frac{\partial L}{\partial y}$ we obtain our result from the Euler-Lagrange equation. We must note here that both the equations (\ref{eqx}) and (\ref{eqy}) contain $\ddot{x}$ and $\ddot{y}$ terms. The equations of motion are therefore obtained by separating these terms.
\begin{prop}
The Euler-Lagrange equation for the relativistic Kapitza Lagrangian yields the relativistic Kapitsa equation
\be\label{kapx}
\ddot{x} = \frac{\dot{x}\dot{y}}{\Gamma c^2} ( -ky + bx) - \frac{1}{\Gamma \Gamma_{x}^{2}} (kx + by), \ee
\be
\ddot{y} = \frac{\dot{x}\dot{y}}{\Gamma c^2} ( kx + by) + \frac{1}{\Gamma \Gamma_{y}^{2}} (-ky + bx), \ee
where
\be\label{kapy}
\Gamma_x = ( 1 -  \frac{\dot{x}^2}{c^2})^{-\frac{1}{2}}, \quad \Gamma_y = ( 1 +  \frac{\dot{y}^2}{c^2})^{- \frac{1}{2}}.
\ee
\end{prop}
{\bf Proof} \, Multiplying (\ref{eqx})  by $( 1 -  \frac{\dot{x}^2}{c^2})$ we obtain
$$
\ddot{x}\big( 1 + \frac{\dot{y}^{2}}{c^2} \big)\big( 1 - \frac{\dot{x}^{2}}{c^2} \big) - \frac{\dot{x}\dot{y}}{c^2}
\ddot{y}\big( 1 - \frac{\dot{x}^{2}}{c^2} \big) = - \big(\frac{kx}{\Gamma^3} + \frac{by}{\Gamma^3}\big)\big( 1 - \frac{\dot{x}^{2}}{c^2} \big)
$$
$$
\frac{\ddot{x}}{\Gamma^2}= \frac{\dot{x}\dot{y}}\Gamma^3{c^2}\big(-ky + bx \big) - \frac{1}{\Gamma_{x}^{2} \Gamma^3}\big( kx + by).
$$
Similarly, we obtain the second equation via the multiplication of Eq.(\ref{eqy}) with  $( 1 +  \frac{\dot{y}^2}{c^2})$.
\begin{cor}
 The equations  (\ref{kapx}) and  (\ref{kapx}) for non-relativistic limit reduce to the Kapitsa equation
\be
	\ddot{x} + kx + by = 0, \qquad \ddot{y} - ky + bx = 0.
\ee
\end{cor}
The relativistic Kapitza equation can also be formulated using Hamiltonian method.
\begin{prop}
The (effective) Hamiltonian of the relativistic Kapitsa equation is
\be
H = \frac{p_{x}^{2}}{2\Gamma} - \frac{p_{y}^{2}}{2\Gamma} + \frac{1}{2}k(x^2 - y^2) + bxy,
\ee
and corresponding Hamiltonian equations are given by
\be\label{Hamx}
\dot{x} = \frac{\partial H}{\partial p_x} = \frac{p_x}{\Gamma}, \quad \dot{p_x} = -\frac{\partial H}{\partial x} = -kx - by,
\ee
\be\label{Hamy}
\dot{y} = \frac{\partial H}{\partial p_y} = -\frac{p_y}{\Gamma}, \quad \dot{p_y} = -\frac{\partial H}{\partial y} = ky -bx.
\ee
\end{prop}
One can clearly see that $p_x = \dot{x}\Gamma$, $p_y = -\dot{y}\Gamma$ and the rest follows from lemma 3.1. It can be easily shown that the relativistic Kapitsa equation follows from Eqns.(\ref{Hamx}) and (\ref{Hamy}).

\bigskip

It is clear that for the non-relativistic limit the Hamiltonian equation yields the Kapitza equation. One must note that
non-relativistic Kapitza equation has two integrals of motion, the Hamiltonian $H = \frac{p_{x}^{2}}{2} - \frac{p_{y}^{2}}{2} + \frac{1}{2}k(x^2 - y^2) + bxy$ and the Fradkin tensor $I = p_xp_y + \frac{1}{2}b(x^2 - y^2) - kxy$. Unfortunately, for the relativistic case we have just one conserved quantity due to the presence of the Lorentz factor.

\subsection{Relativistic rotation of saddle}

The Lagrangian of the rotating saddle potential is given by
\be
L_{\text{rel saddle}} = -\frac{c^2}{\Gamma} - \frac{\Lambda}{2}
\Big( \big(x^2 - y^2 \big)\cos(2\Gamma\omega t)
 + 2xy \sin(2\Gamma\omega t) \Big),
 \ee
where the constant coefficients $b$ and $k$ are replaced by the sinusoidal terms,
the corresponding equations of motion are
\be\label{kapx}
\ddot{x} = \frac{\dot{x}\dot{y}}{\Gamma c^2} \Lambda \big(-\cos(2\Gamma\omega t) y + \sin(2\Gamma\omega t) x \big) -
\frac{1}{\Gamma \Gamma_{x}^{2}} \Lambda\big(\cos(2\Gamma\omega t)x + \sin(2\Gamma\omega t)y \big), \ee
\be\label{kapx2}
\ddot{y} = \frac{\dot{x}\dot{y}}{\Gamma c^2}\Lambda \big( \cos(2\Gamma\omega t)x + \sin(2\Gamma\omega t)y \big)
+ \frac{1}{\Gamma \Gamma_{y}^{2}} \Lambda \big(-\cos(2\Gamma\omega t)y + \sin(2\Gamma\omega t)x \big), \ee
where
\be\label{kapy}
\Gamma_x = ( 1 -  \frac{\dot{x}^2}{c^2})^{-\frac{1}{2}}, \quad \Gamma_y = ( 1 +  \frac{\dot{y}^2}{c^2})^{- \frac{1}{2}}.
\ee
%==============================
\begin{figure}
\begin{subfigure}{.5\textwidth}
  \centering
  % include first image
  \includegraphics[height=5 cm,width=5.9 cm]{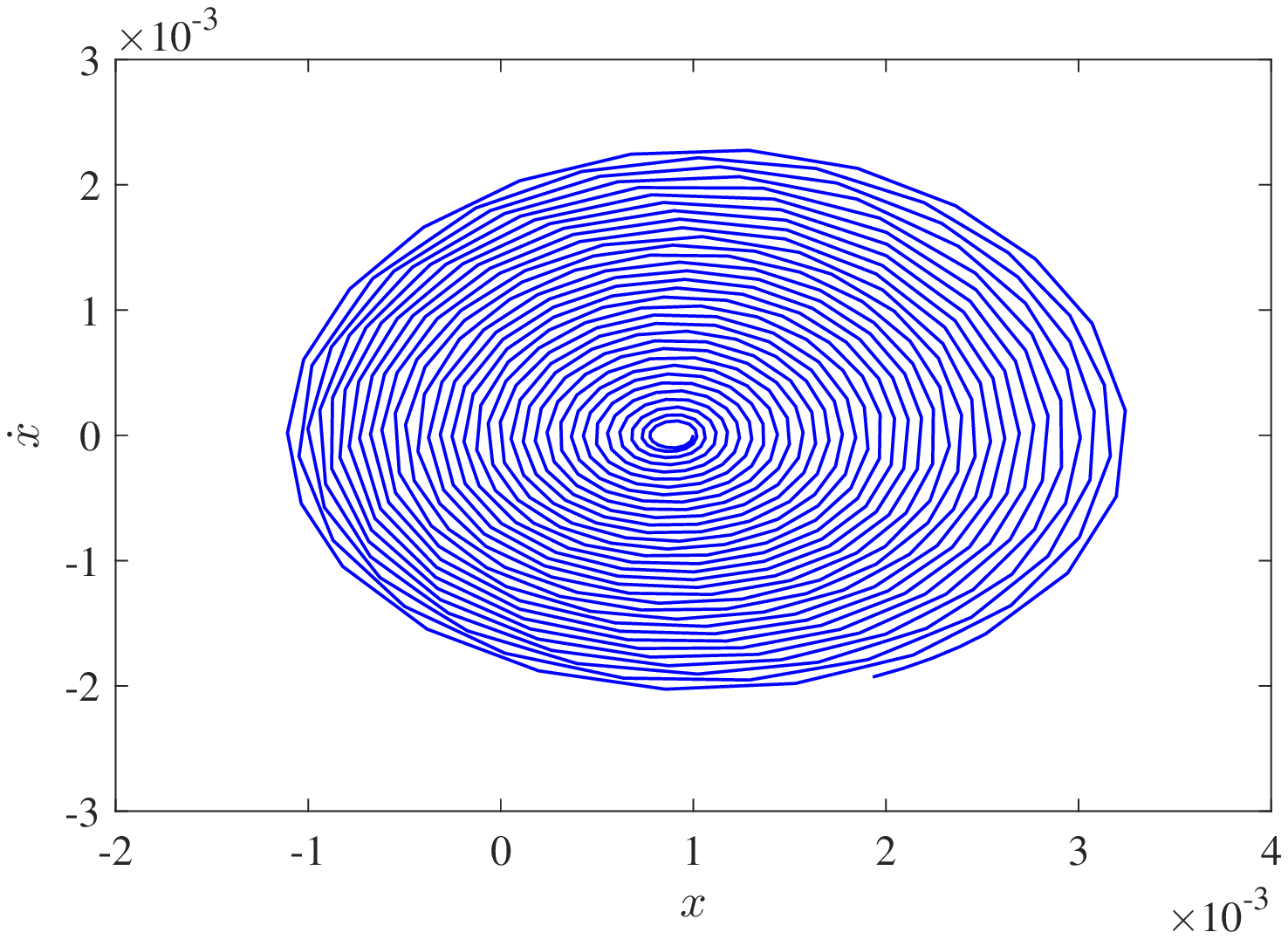}
  \caption{Plot of $\dot{x}$ vs $x$.}
  \label{f1a}
\end{subfigure}
\begin{subfigure}{.5\textwidth}
  \centering
  % include second image
  \includegraphics[height=5 cm,width=5.9 cm]{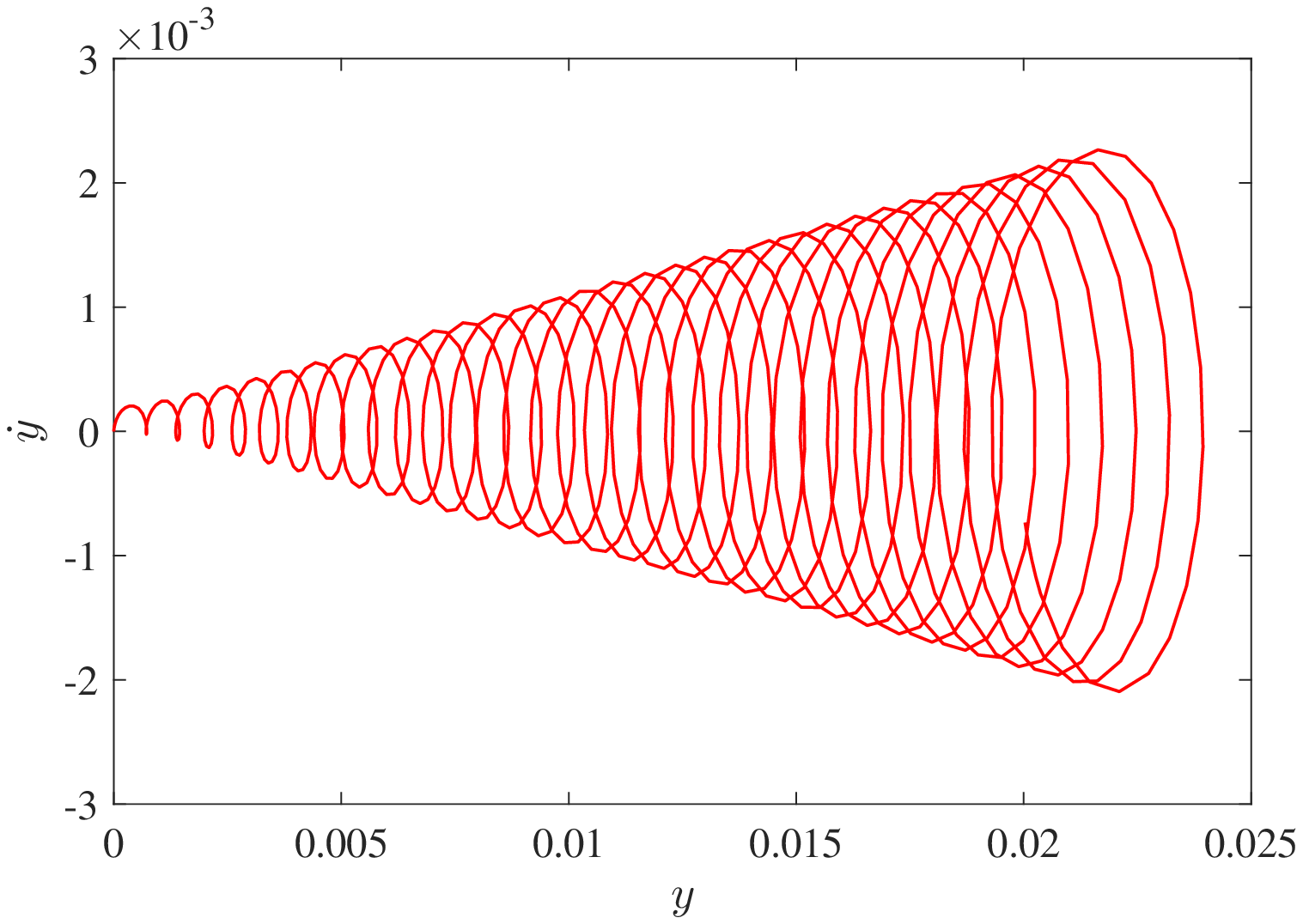}
  \caption{Plot of $\dot{y}$ vs $y$.}
  \label{f1b}
\end{subfigure}

%\newline

\begin{subfigure}{.5\textwidth}
  \centering
  % include third image
  \includegraphics[height=5 cm,width=5.9 cm]{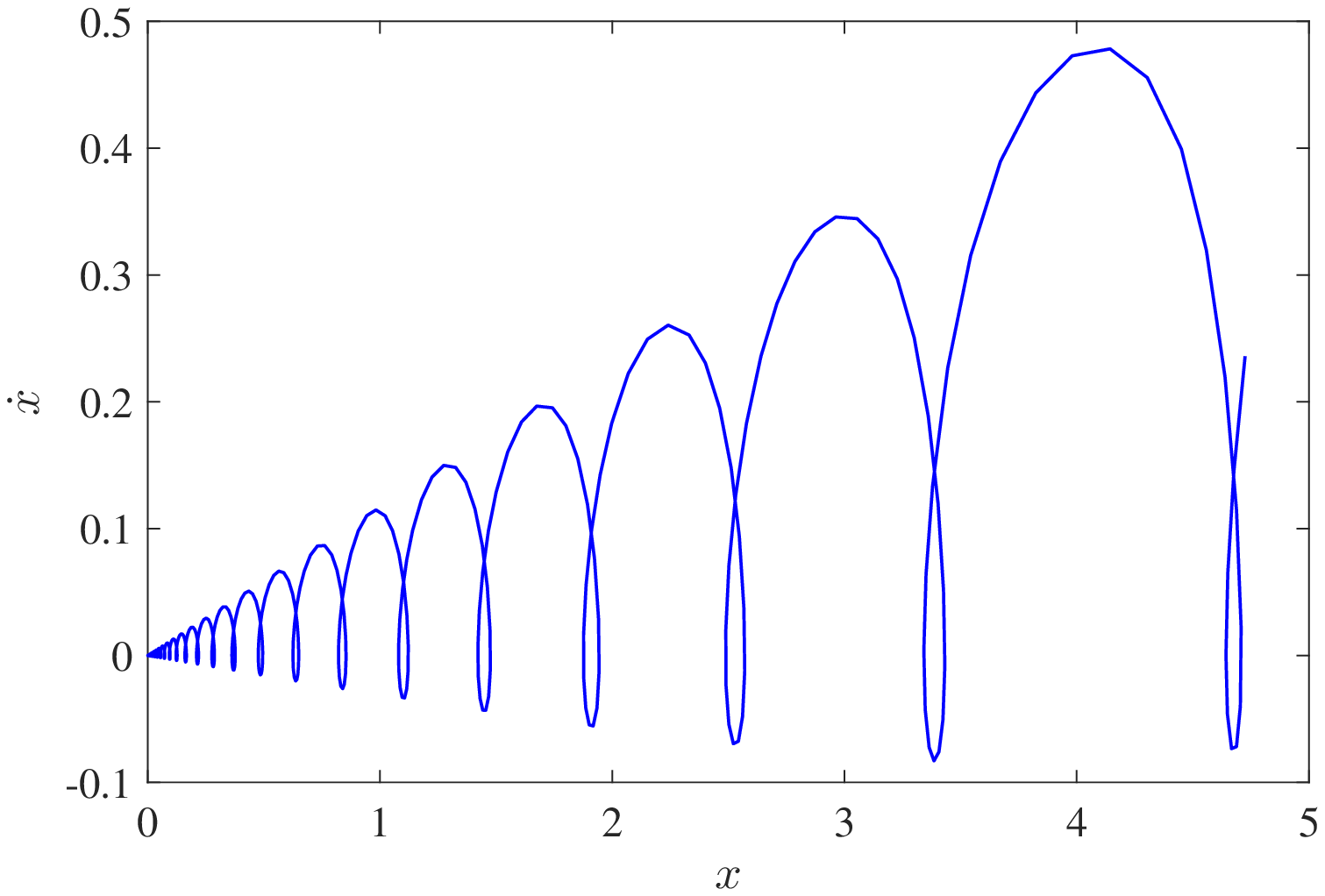}
  \caption{Plot of $\dot{x}$ vs $x$.}
  \label{f1c}
\end{subfigure}
\begin{subfigure}{.5\textwidth}
  \centering
  % include fourth image
  \includegraphics[height=5 cm,width=5.9 cm]{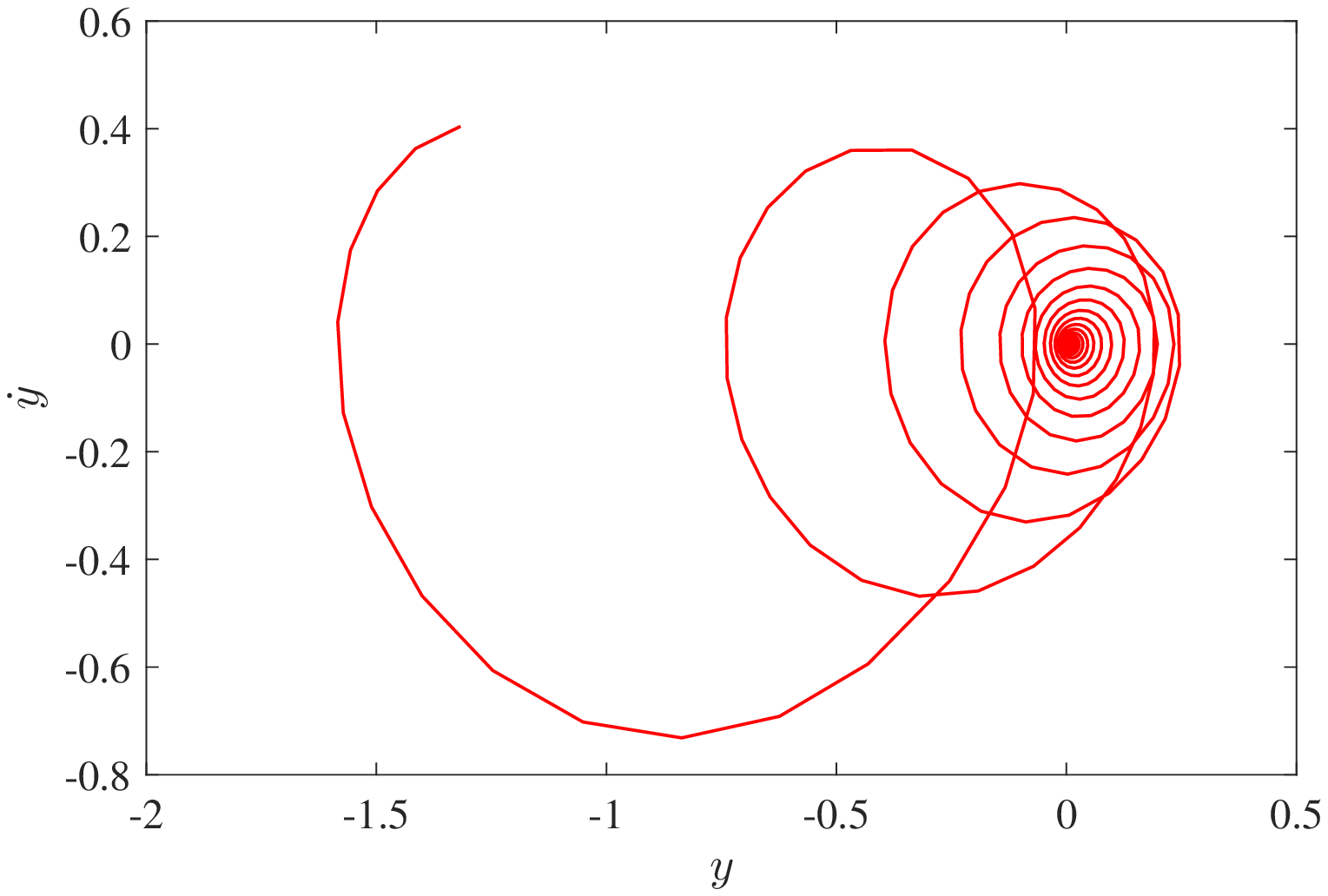}
  \caption{Plot of $\dot{y}$ vs $y$.}
  \label{f1d}
\end{subfigure}
\caption{Phase space diagrams (with normalized parameters) via Eq.(\ref{kapx}) and Eq.(\ref{kapx2}) with  $\Lambda= 0.1$ and $\omega=0.5$. The top two figures are for non-relativistic cases with $v_x=0=v_y$ and the bottom two are for the relativistic cases with $v_x=0.5c=v_y$. The initial conditions are $x(0)=0$, $\dot{x}(0)=0.0$, $y(0)=0.0$, $\dot{y}(0)=0.001$. The normalized parameters are as follows: $x \rightarrow x \omega_c /c$, $y \rightarrow y \omega_c /c$, $\dot{x} \rightarrow \dot{x}/c$, $\dot{y} \rightarrow \dot{y}/c$, $\ddot{x} \rightarrow \ddot{x} \omega_c c$, $\ddot{y} \rightarrow \ddot{y} \omega_c c$, $\Lambda \rightarrow \Lambda/\omega_c ^2$, $\omega \rightarrow \omega/\omega_c$ and $t \rightarrow t \omega_c$ where $\omega_c$ can be taken as the cyclotron frequency in the typical microwave range.}
\label{f1}
\end{figure}
%=================================================================
%
The phase space plots for the rotating saddle have been depicted in figure (\ref{f1}) for Eqs.(\ref{kapx}) \& (\ref{kapx2}). In the non-relativistic scenario, it can be seen from figure (\ref{f1a}) that the trajectory corresponds to limit cycle under one specific saddle point (SP) and spanning over smaller to larger amplitudes. The particle can be trapped inside those periodic boundaries for shorter times but not completely. Figure (\ref{f1b}) corresponds to heteroclinic orbit where it moves from one SP to another over time directs to the escaping of particles. In the relativistic case, both figure (\ref{f1c}) and (\ref{f1d}) correspond heteroclinic orbit and particle escaping.

\bigskip

In the non-relativistic limit Eqns.(\ref{kapx}) and (\ref{kapy}) reduce to
$$
\ddot{x} + \Lambda x\cos(2\omega t) -  \Lambda y \sin(2\omega t) = 0, \qquad
\ddot{y} + \Lambda y\cos(2\omega t) - \Lambda x \sin(2\omega t) = 0.
$$
The relativistic generalized Kapitza equation corresponding monkey saddle is the extension of the monkey saddle equation,
\be\label{grkapx}
\ddot{x} = \frac{\dot{x}\dot{y}}{\Gamma c^2} \big(-2k_1 xy + k_2 (x^2 - y^2) \big) -
\frac{1}{\Gamma \Gamma_{x}^{2}}\big(k_1(x^2 - y^2) + 2k_2 xy \big),
\ee
\be\label{grkapy}
\ddot{y} = \frac{\dot{x}\dot{y}}{\Gamma c^2} \big(k_1(x^2 - y^2) + 2 k_2 xy \big)  +
\frac{1}{\Gamma \Gamma_{y}^{2}}\big(-2k_1 xy + k_2 (x^2 - y^2) \big).
\ee
%
%==============================
\begin{figure}
\begin{subfigure}{.5\textwidth}
  \centering
  % include first image
  \includegraphics[height=5 cm,width=5.9 cm]{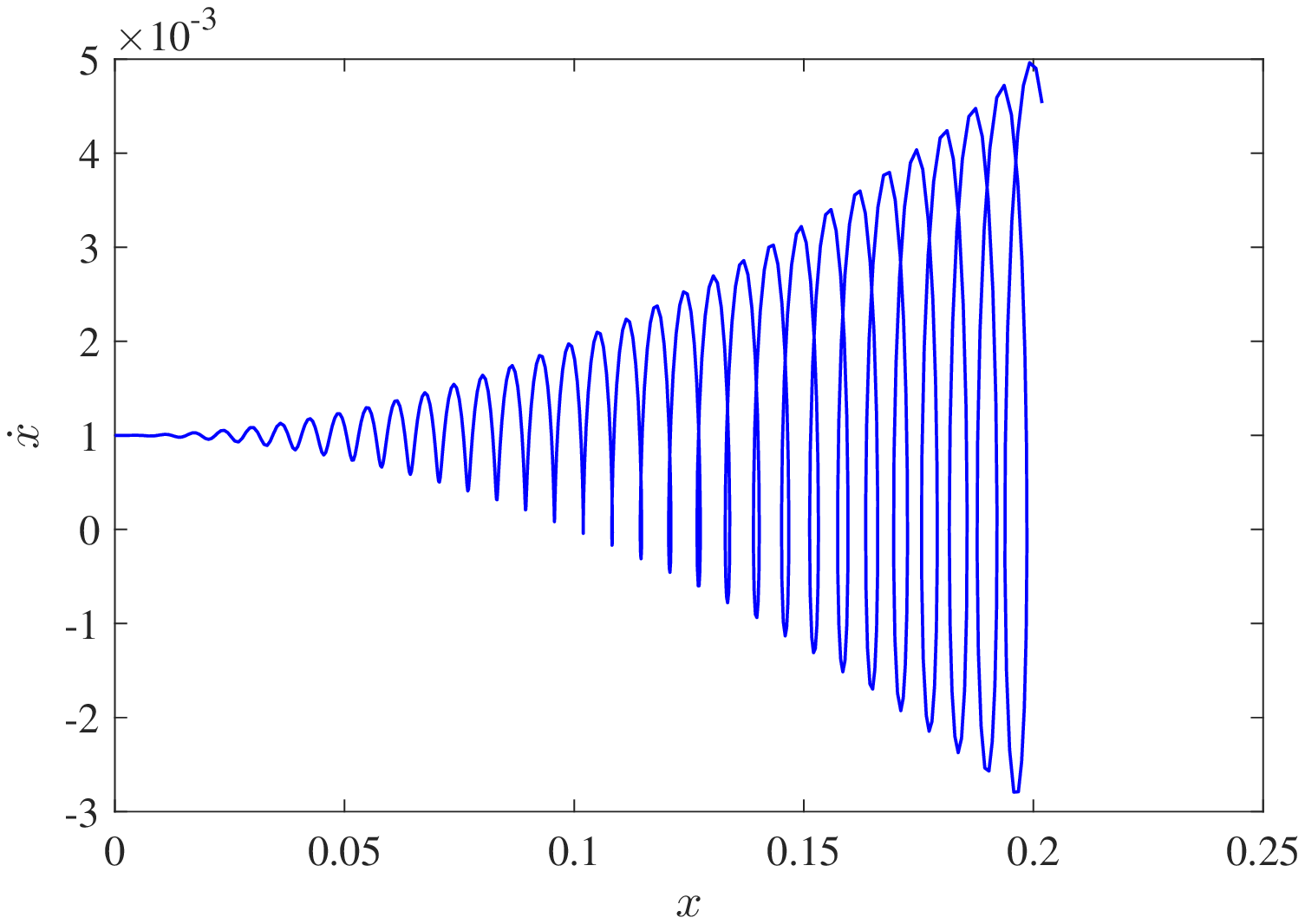}
  \caption{Plot of $\dot{x}$ vs $x$.}
  \label{f2a}
\end{subfigure}
\begin{subfigure}{.5\textwidth}
  \centering
  % include second image
  \includegraphics[height=5 cm,width=5.9 cm]{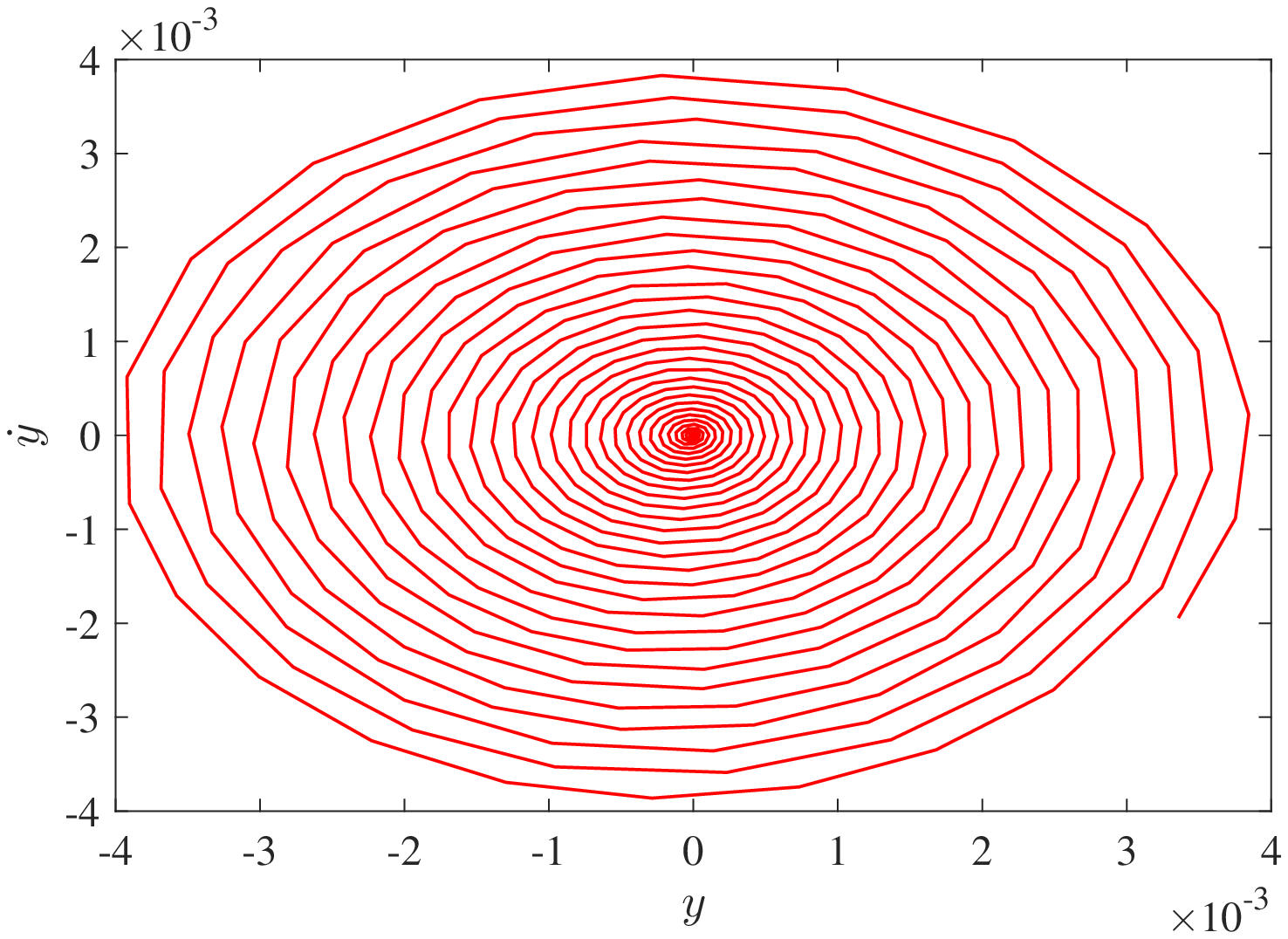}
  \caption{Plot of $\dot{y}$ vs $y$.}
  \label{f2b}
\end{subfigure}

%\newline

\begin{subfigure}{.5\textwidth}
  \centering
  % include third image
  \includegraphics[height=5 cm,width=5.9 cm]{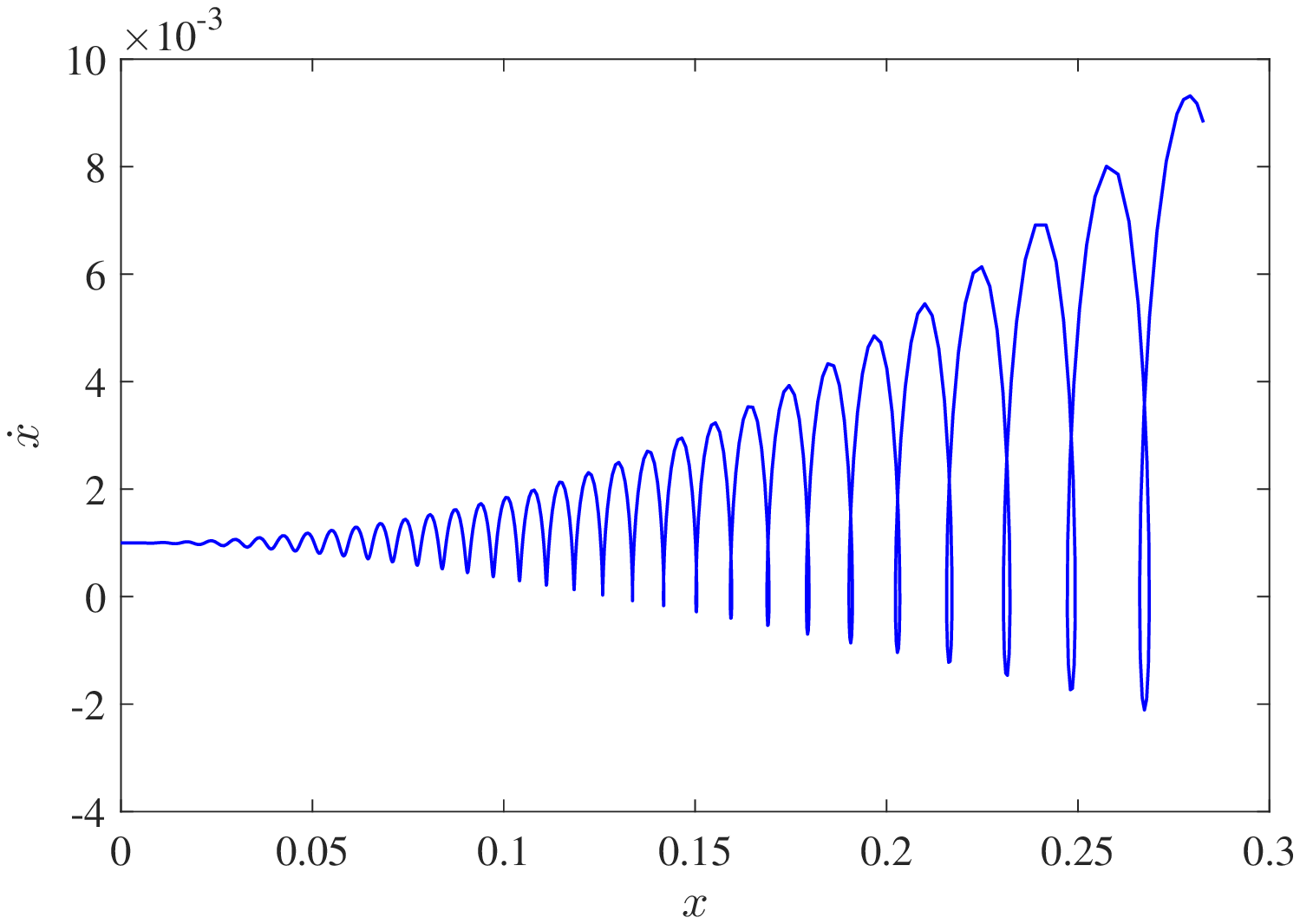}
  \caption{Plot of $\dot{x}$ vs $x$.}
  \label{f2c}
\end{subfigure}
\begin{subfigure}{.5\textwidth}
  \centering
  % include fourth image
  \includegraphics[height=5 cm,width=5.9 cm]{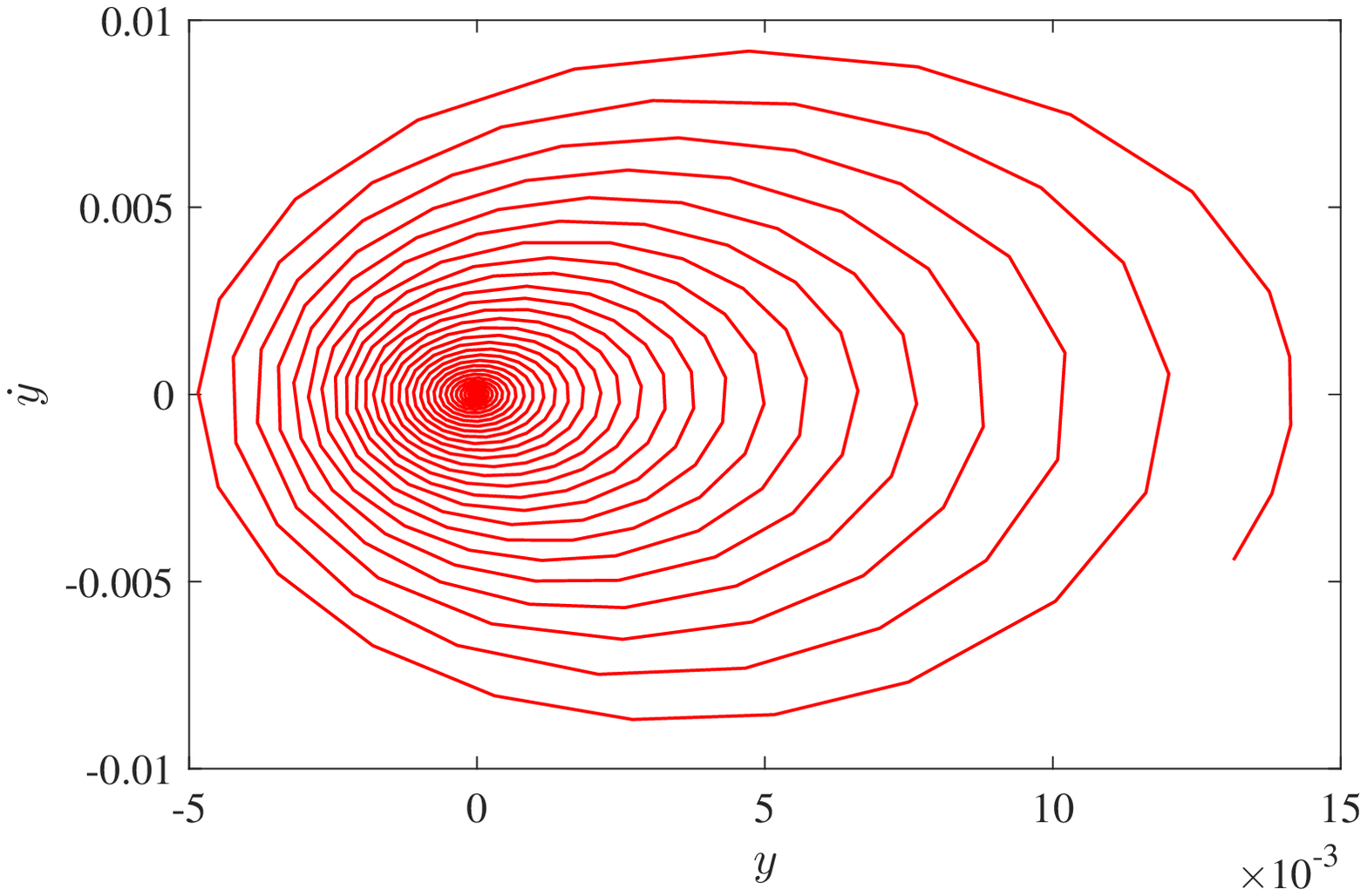}
  \caption{Plot of $\dot{y}$ vs $y$.}
  \label{f2d}
\end{subfigure}
\caption{Phase space diagrams (with normalized parameters) via Eq.(\ref{grkapx}) and Eq.(\ref{grkapy}) with  $\Lambda= 0.1$, $\omega=0.5$ and $\alpha_1 = \alpha_2 = 1$. The top two figures are for non-relativistic cases with $v_x=0=v_y$ and the bottom two are for the relativistic cases with $v_x=0.5c=v_y$. The initial conditions are $x(0)=0$, $\dot{x}(0)=0.0$, $y(0)=0.0$, $\dot{y}(0)=0.001$. The normalization scheme is same as mentioned in figure (\ref{f1}) along with  $\alpha_1 \rightarrow \alpha_1 c/ \omega_c$ and $\alpha_2 \rightarrow \alpha_2 c/ \omega_c$.}
\label{f2}
\end{figure}
%=================================================================
%
The phase space plots for the relativistic rotating monkey saddle have been depicted in figure (\ref{f2}) for Eqs.(\ref{grkapx}) \& (\ref{grkapy}). Figure (\ref{f2a}) and (\ref{f2c}) show heteroclinic trajectories which remove the possibility of particle trapping for both relativistic and non-relativistic cases. In contrary to this figure (\ref{f2b}) and (\ref{f2d}) show limit cycle trajectories under one SP for both relativistic and non-relativistic cases. As one can see that there are periodic boundaries which are changing very slowly over time and closely spaced, henceforth particle can be trapper over shorter time scale.

The {\it relativistic rotating monkey saddle equation} can be obtained by replacing $k_1 = \Lambda \alpha_1 \cos(\Gamma 2\omega t)$
and $k_2 = \Lambda \alpha_2 \sin(\Gamma 2\omega t)$.

\subsection{Flapping and spinning of relativistic saddle potential and trap}

Suppose we allow the potential $U = g(\frac{1}{2}(x^2 - y^2))$ to vary  sinusoidally with time. this becomes
\be
{\cal U}(x,y,t) = A_{RF}\cos(\omega t)g(\frac{1}{2}(x^2 - y^2),
\ee
%----------------------------------------------------------------------------------------
where $ A_{RF}$ is the amplitude of the AC component of the electric field applied to create the trap, and $\omega$ is the frequency. This can be viewed as a {\it nonlinear} generalization of `flapping saddle'. Latter occurs in a quadrupole potential in which extreme oscillate between peaks and valleys. Thus Newton's law yields the following pair of equations
\be
\ddot{x} = -A_{RF} \cos(\omega t) xg(\frac{1}{2}(x^2 - y^2), \quad  \ddot{y} = -A_{RF} y\cos(\omega t) g(\frac{1}{2}(x^2 - y^2).
\ee
%
%==============================
\begin{figure}[ht]
\begin{subfigure}{.5\textwidth}
  \centering
  % include first image
  \includegraphics[height=5 cm,width=5.9 cm]{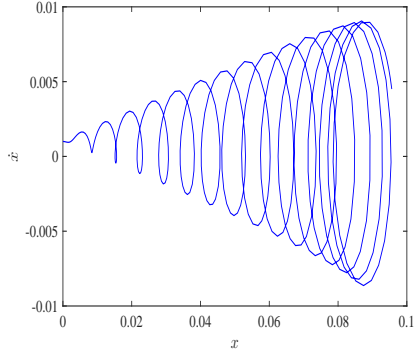}
  \caption{Plot of $\dot{x}$ vs $x$.}
  \label{f3a}
\end{subfigure}
\begin{subfigure}{.5\textwidth}
  \centering
  % include second image
  \includegraphics[height=5 cm,width=5.9 cm]{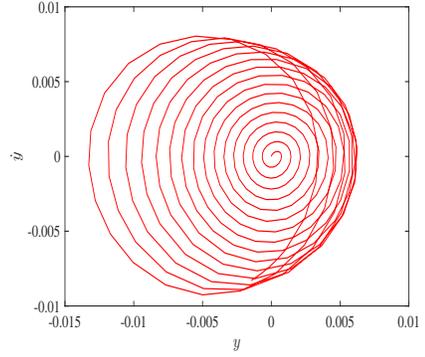}
  \caption{Plot of $\dot{y}$ vs $y$.}
  \label{f3b}
\end{subfigure}

%\newline

\begin{subfigure}{.5\textwidth}
  \centering
  % include third image
  \includegraphics[height=5 cm,width=5.9 cm]{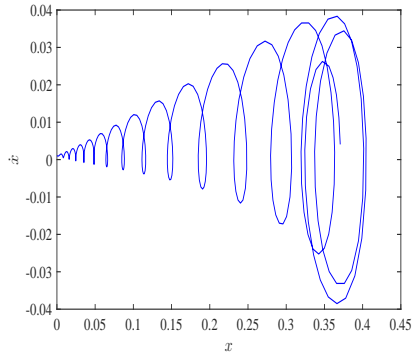}
  \caption{Plot of $\dot{x}$ vs $x$.}
  \label{f3c}
\end{subfigure}
\begin{subfigure}{.5\textwidth}
  \centering
  % include fourth image
  \includegraphics[height=5 cm,width=5.9 cm]{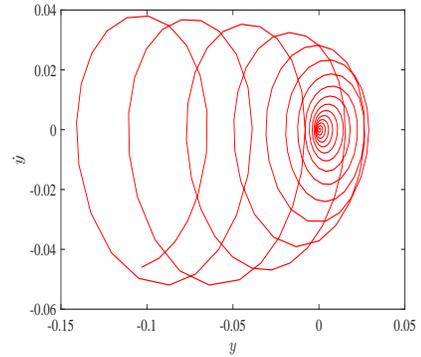}
  \caption{Plot of $\dot{y}$ vs $y$.}
  \label{f3d}
\end{subfigure}
\caption{Phase space diagrams (with normalized parameters) via Eq.(\ref{rcdex}) and Eq.(\ref{rcdey}) with  $\Lambda= 0.1$, $\omega=0.5$ and $\beta_1=\beta_2=\beta_3=\beta_4=1$. The top two figures are for non-relativistic cases with $v_x=0=v_y$ and the bottom two are for the relativistic cases with $v_x=0.5c=v_y$. The initial conditions are $x(0)=0$, $\dot{x}(0)=0.0$, $y(0)=0.0$, $\dot{y}(0)=0.001$. The normalization scheme is same as mentioned in figure (\ref{f1}) along with  $\beta_1 \rightarrow \beta_1 \omega_c ^2/ c^2$, $\beta_2 \rightarrow \beta_2 \omega_c ^4/ c^4$, $\beta_3 \rightarrow \beta_3 \omega_c ^2/ c^2$ and $\beta_4 \rightarrow \beta_4 \omega_c ^4/ c^4$. It must be noted here that we have chosen $g(x^2-y^2)=1+\beta_1 (x^2 - y^2) + \beta_2 (x^2 - y^2)^2$ and $f(xy)=1+\beta_3 (xy) + \beta_4 (xy)^2.$ }
\label{f3}
\end{figure}
%=================================================================
%
The linear case coincides with the form of the Mathieu differential equations. Our motivation is to study this new set of equations. In general it is difficult to create flapping saddle on which the ball can move. So we bang on the time-dependent spinning saddle at angular frequency $\omega$, which is similar to the flapping potential
\be
\hat{{\cal U}}(x,y,t) = A\big( g(x^2 - y^2)\cos (2\omega t) + 2f(xy) \sin (2\omega t) \big),
\ee
where $A$ is some constant. We now apply again the relativistic Lagrangian as
\be
L_{relflap} = - \frac{c^2}{\Gamma} - \hat{{\cal U}}(x,y,t),  \quad \hbox{ where } \hat{{\cal U}}(x,y,t)= A\big( g(x^2 - y^2)\cos (2\omega t) + 2f(xy) \sin (2\omega t) \big),
\ee
to get the relativistic coupled differential equations as
\begin{eqnarray}\label{rcdex}
\ddot{x} = \frac{\dot{x}\dot{y}}{\Gamma c^2}\Lambda \big( - y g^{\prime}(x^2 - y^2) \cos (2\Gamma \omega t)
+ x f^{\prime}(xy) \sin(2\Gamma \omega t) \big) \nonumber \\
- \frac{1}{\Gamma \Gamma_{x}^{2}}
\big(x g^{\prime}(x^2 - y^2) \cos (2\Gamma \omega t) + yf^{\prime}(xy) \sin(2\Gamma \omega t) \big),
\end{eqnarray}
\begin{eqnarray}\label{rcdey}
\ddot{y} = \frac{\dot{x}\dot{y}}{\Gamma c^2}\Lambda \big(x g^{\prime}(x^2 - y^2) \cos (2\Gamma \omega t)
+ yf^{\prime}(xy) \sin(2\Gamma \omega t) \big) \nonumber \\
+ \frac{1}{\Gamma \Gamma_{y}^{2}}
\big( - y g^{\prime}(x^2 - y^2) \cos (2\Gamma \omega t)
+ x f^{\prime}(xy) \sin(2\Gamma \omega t) \big).
\end{eqnarray}
If we take $\tau = \omega t$ and $x^{\prime} = \frac{dx}{d\tau}$ we obtain
\begin{eqnarray}
x^{\prime \prime} =  \frac{x^{\prime} y^{\prime}}{\Gamma c^2}\Lambda \big( -
y g^{\prime}(x^2 - y^2) \cos (2\Gamma \omega t)
+ x f^{\prime}(xy) \sin(2\Gamma \omega t) \big) \nonumber \\
 - \frac{1}{\omega^2\Gamma \Gamma_{x}^{2}}
\big(x g^{\prime}(x^2 - y^2) \cos (2\Gamma \omega t) + yf^{\prime}(xy) \sin(2\Gamma \omega t) \big),
\end{eqnarray}
\begin{eqnarray}
y^{\prime \prime} =  \frac{x^{\prime} y^{\prime}}{\Gamma c^2}\Lambda
\big(x g^{\prime}(x^2 - y^2) \cos (2\Gamma \omega t)
+ yf^{\prime}(xy) \sin(2\Gamma \omega t) \big) \nonumber \\
+ \frac{1}{\omega^2\Gamma \Gamma_{y}^{2}}
\big( - y g^{\prime}(x^2 - y^2) \cos (2\Gamma \omega t)
+ x f^{\prime}(xy) \sin(2\Gamma \omega t) \big).
\end{eqnarray}
The phase space plots for the relativistic flapping saddle have been depicted in figure (\ref{f3}) for Eqs.(\ref{rcdex}) \& (\ref{rcdey}). Like in the previous case of relativistic rotating monkey saddle here also we see that both trapping in shorter time scale and escaping is possible in both relativistic and non-relativistic limits.

\section{Summary}

The relativistic analog of the Kapitza equation has been investigated  and associated trapping phenomena of the charged particles have been discussed. We have presented an example of a relativistic curl force and the generalization of the curl force associated with saddles potentials formulated by Berry and Shukla. In case of relativistic, rotating saddle the phase space plots correspond some heteroclinic orbits and particle escaping. With the inclusion of the monkey saddle to the generalized Kapitza equation in relativistic domain the phase plots also show heteroclinic trajectories with a possibility of particle trapping in specific parametric domain in the shorter time scale. In case of relativistic flapping and spinning saddle there is also a possibility of trapping and escaping same as discussed in the previous one. At last we have discussed about the relativistic generalization of the Kapitza equation associated with the monkey saddle. The theoretical outcomes  along with the numerical results may direct and or give an idea about the possible mechanism of particle trapping with finite curl forces in the relativistic domain. In the earlier observations\cite{GaraiGuha,GaraiGuha2}, we see that there is a possibility of trapping the charged particles under the curl force dynamics with saddle potentials completely, whereas, in this work we have observed that the trapping is not complete in the relativistic domain. As the results in the relativistic domain direct us of incomplete trapping in a shorter time scale therefore it becomes a field of relative interest to further investigate this with the choice of different saddle potentials and to apply the outcomes towards astrophysical scenarios.

\section*{Acknowledgments}

PG is immensely grateful to \emph{Professor Sir Michael Berry} for his critical and meticulous reading of this draft and for making invaluable remarks and or comments. PG and SG also convey their sincere thanks to \emph{Professor Haret Rosu} for his suggestions and comments and \emph{Professors Praghya Shukla, Stefan Mancas and Anindya Ghose-Choudhury} for various discussions and correspondences. PG thanks \emph{Khalifa University of Science and Technology} for its continued support towards this research work under the grant number FSU-2021-014 and SG thanks \emph{Diamond Harbour Women’s University} for providing the necessary research environment with constant encouragement and support.

\section*{Author contributions}

All the authors contributed equally to this work

\section*{Conflicts of interest}

The authors declare that they have no conflict of interest.

\section*{Data transparency}

Not applicable.

\end{document}